\documentclass[twocolumn,showpacs,preprintnumbers,superscriptaddress,amsmath,floatfix,amssymb,secnumarabic,prl]{revtex4}

\usepackage{amscd}
\usepackage{amssymb}
\usepackage{amsmath}
\usepackage{color}
\usepackage{placeins}

\usepackage[colorlinks=true]{hyperref}
\usepackage{graphicx}
\graphicspath{{./plots/}}
\usepackage[sort&compress]{natbib}
\bibpunct{[}{]}{,}{n}{}{}

\usepackage{etoolbox}
\newcommand{\mysection}[2]{\textbf{#1.}\,\,}
\newtoggle{arxiv}
\togglefalse{arxiv} 

\definecolor{light-gray}{gray}{0.75}
\newtoggle{draft}
\toggletrue{draft}

\DeclareMathOperator{\arccosh}{arccosh}

\newcommand{\comment}[1]{}

\newcommand{\lr}[1]{ \left( #1 \right) }
\newcommand{\lrs}[1]{ \left[ #1 \right] }

\newcommand{\Tr}{ {\rm Tr} \, }
\newcommand{\tr}{ {\rm tr} \, }
\newcommand{\re}{ {\rm Re} \, }
\newcommand{\im}{ {\rm Im} \, }

\newcommand{\floor}[1]{\lfloor #1 \rfloor}

\renewcommand{\det}[1]{ {\rm det} \left( #1 \right) }

\newcommand{\expa}[1]{ \exp{\left( #1 \right)} }

\newcommand{\Z}{\mathcal{Z}}

\renewcommand{\d}{{\rm d} \/}

\newcommand{\mpipi}{\left[-\pi,\right.\left. \pi\right)}

\begin{document}
\sloppy

\title{Complex Saddles in Two-dimensional Gauge Theory}

\author{P.~V.~Buividovich}
\email{pavel.buividovich@physik.uni-regensburg.de}
\affiliation{Institute for Theoretical Physics, Regensburg University, D-93053 Regensburg, Germany}

\author{Gerald~V.~Dunne}
\email{dunne@phys.uconn.edu}
\affiliation{Department of Physics, University of Connecticut, Storrs, CT 06269-3046, USA}

\author{S.~N.~Valgushev}
\email{semen.valgushev@physik.uni-regensburg.de}
\affiliation{Institute for Theoretical Physics, Regensburg University, D-93053 Regensburg, Germany}
\affiliation{Institute of Theoretical and Experimental Physics (ITEP), 117218 Russia, Moscow, B. Cheremushkinskaya str. 25}

\date{\today}
\begin{abstract}
 We study numerically the saddle point structure of two-dimensional (2D) lattice gauge theory, represented by the Gross-Witten-Wadia unitary matrix model. The saddle points are in general complex-valued, even though the original integration variables and action are real. We confirm the trans-series/instanton gas structure in the weak-coupling phase, and identify a new complex-saddle interpretation of non-perturbative effects in the strong-coupling phase. In both phases, eigenvalue tunneling refers to eigenvalues moving off the real interval, into the complex plane, and the weak-to-strong coupling phase transition is driven by saddle condensation.
\end{abstract}
\maketitle

\noindent{\bf Introduction:} Path integral saddle points  are physically important in quantum mechanics, matrix models, quantum field theory (QFT) and string theory, and are deeply related to the typical asymptotic nature of  weak coupling perturbative expansions. Such relations are central to the concept of {\it resurgence}, whereby different saddles are intertwined by monodromy properties that connect them and account for Stokes phases. The theory of resurgence has recently provided  new insights into matrix models and string theories \cite{Marino:07:1,Marino:08:1,Schiappa:10:1,Marino:12:1,Schiappa:12:1,Schiappa:15:1}, and has been applied to asymptotically free QFTs and sigma models \cite{Unsal:12:2,Dunne:12:1,Unsal:14:1,Unsal:15:1}, and localizable SUSY QFTs \cite{Schiappa:15:2}. One motivation for such QFT studies is to find a practical numerical implementation of a semiclassical expansion that could provide a Picard-Lefschetz thimble decomposition of gauge theory, either in the continuum or on the lattice, especially for theories with a sign problem \cite{Scorzato:12:1,Kikukawa:13:1}. A unifying theme in these studies, and in related work \cite{David:93:1,Witten:11:1,Witten:11:2,Dunne:11:2}, is the appreciation that {\it complex} saddles are important, in particular in the context of phase transitions, even though the original `path integral' may be a sum over only real configurations.

 In gauge theories, there are two physical parameters which control the strength of fluctuations around the saddle points and enter the resurgent trans-series expansion: the rank $N$ of the gauge group and the t'Hooft coupling $\lambda\equiv N g^2$, with gauge coupling $g^2$ \cite{MarinoInstantonsLargeNBook}. The interplay between the dependence on $N$ and $\lambda$ leads to novel effects \cite{Marino:07:1,Marino:08:1,Schiappa:15:1} which we explore here. An important goal would be to construct uniform resurgent approximations \cite{Dunne:15:1} (with respect to $\lambda$ and $1/N$) which analytically relate the weak- and strong-coupling phases. For gauge theories, such a relation would certainly improve our understanding of confinement and dynamical mass gap generation. It would also extend the applicability of Diagrammatic Monte-Carlo studies of non-Abelian lattice gauge theories, which are so far limited to the regime of unphysically strong bare coupling constants \cite{deForcrand:14:1}.

 The difference between weak- and strong-coupling phases is particularly dramatic in the large-$N$ limit of 2D gauge theories, where they are separated by a third-order phase transition with respect to the t'Hooft coupling  $\lambda$ \cite{Gross:80:1,Wadia:12:1,Wadia:80:1,Neuberger:81:1} and/or the manifold area $A$ \cite{Douglas:93:2,Matytsin:94:1}. Physically, on the weak-coupling side this large-$N$ phase transition in 2D gauge theory is related to the condensation of instantons \cite{Neuberger:80:1,Neuberger:81:1,Matytsin:94:1}, which are exponentially suppressed at large $N$ away from the transition point. Much less is known about the role of instantons (or other saddles) on the strong-coupling side of this transition, except in the double-scaling limit. Here we study the simplest example of 2D lattice gauge theory, the Gross-Witten-Wadia unitary matrix model \cite{Gross:80:1,Wadia:12:1,Wadia:80:1}, to demonstrate the novel properties of complex saddles in the strong-coupling phase as well as their relation to the resurgent structure of the $1/N$ expansion.

\begin{figure*}[h!tpb]
\includegraphics[width=0.17\linewidth]{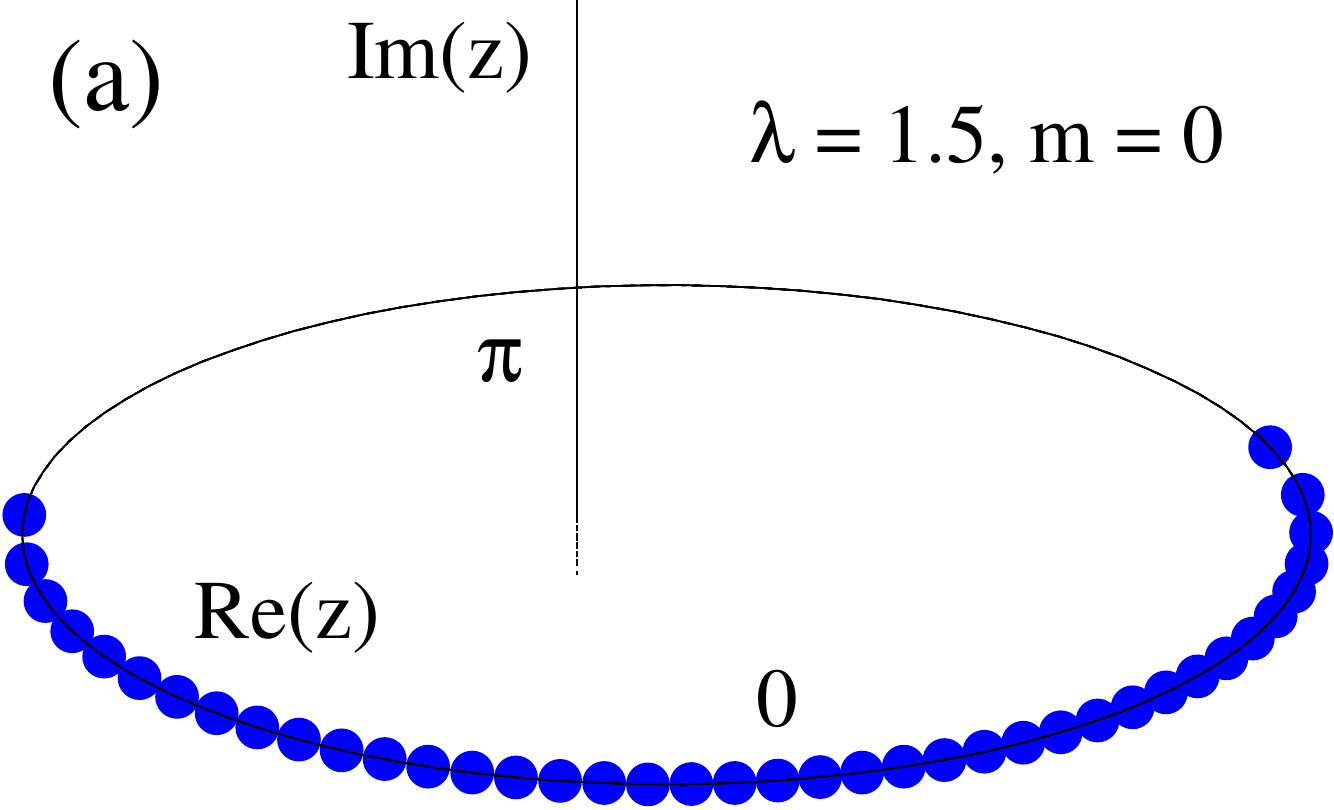}
\includegraphics[width=0.17\linewidth]{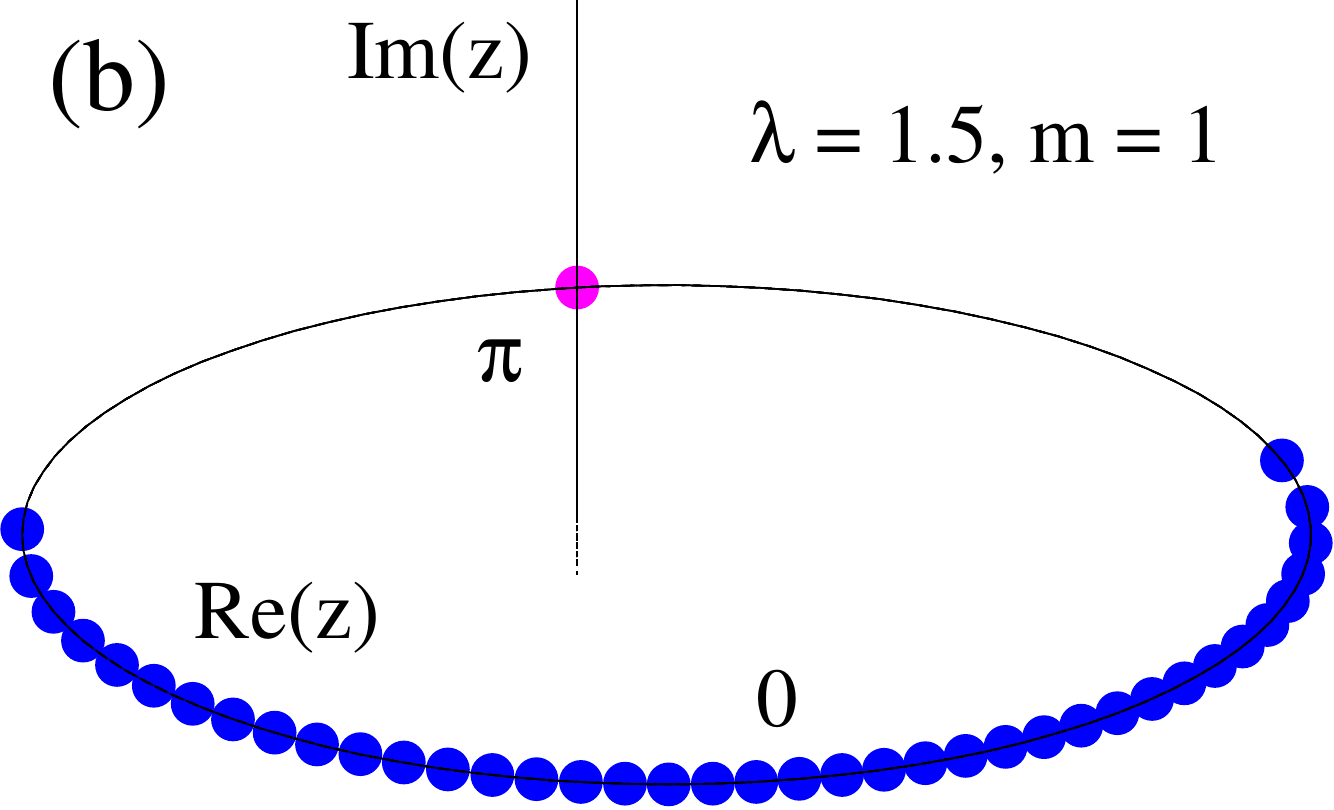}
\includegraphics[width=0.17\linewidth]{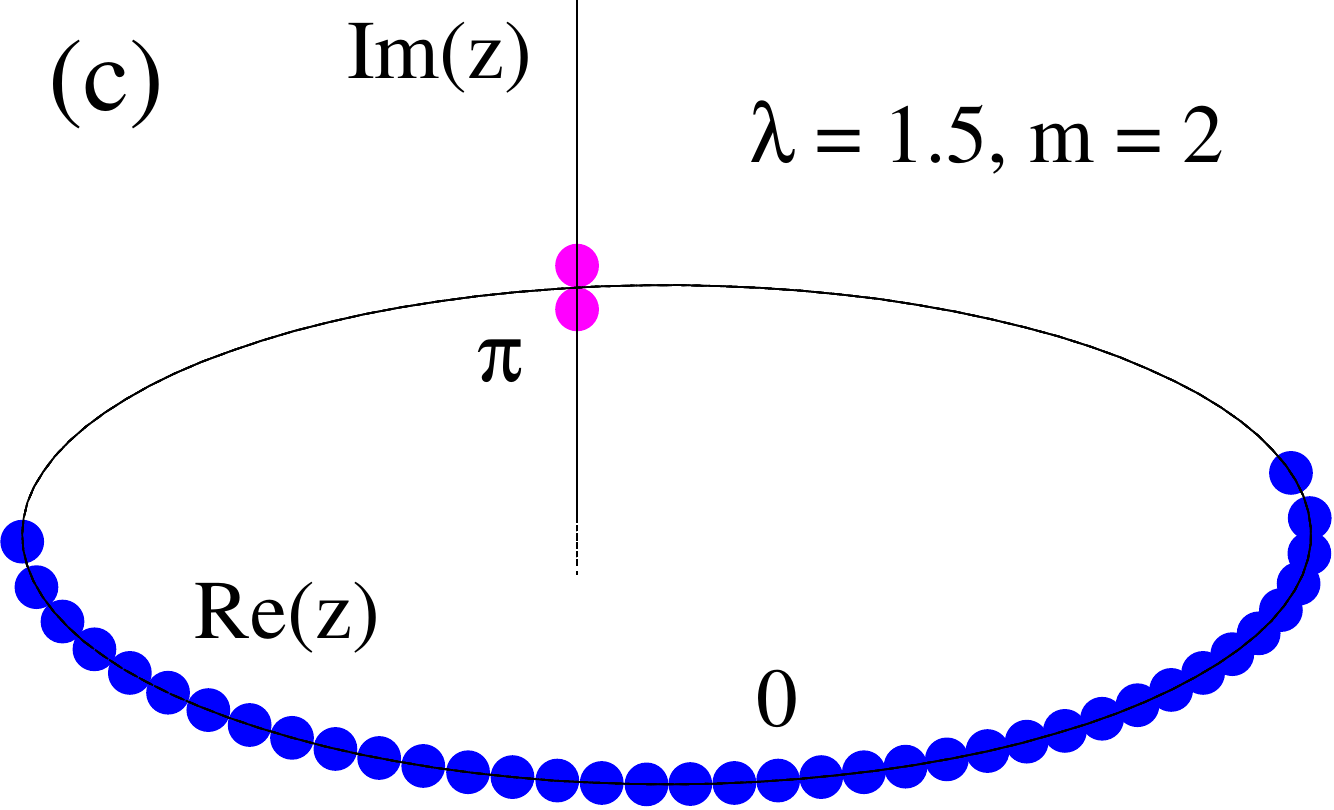}
\includegraphics[width=0.17\linewidth]{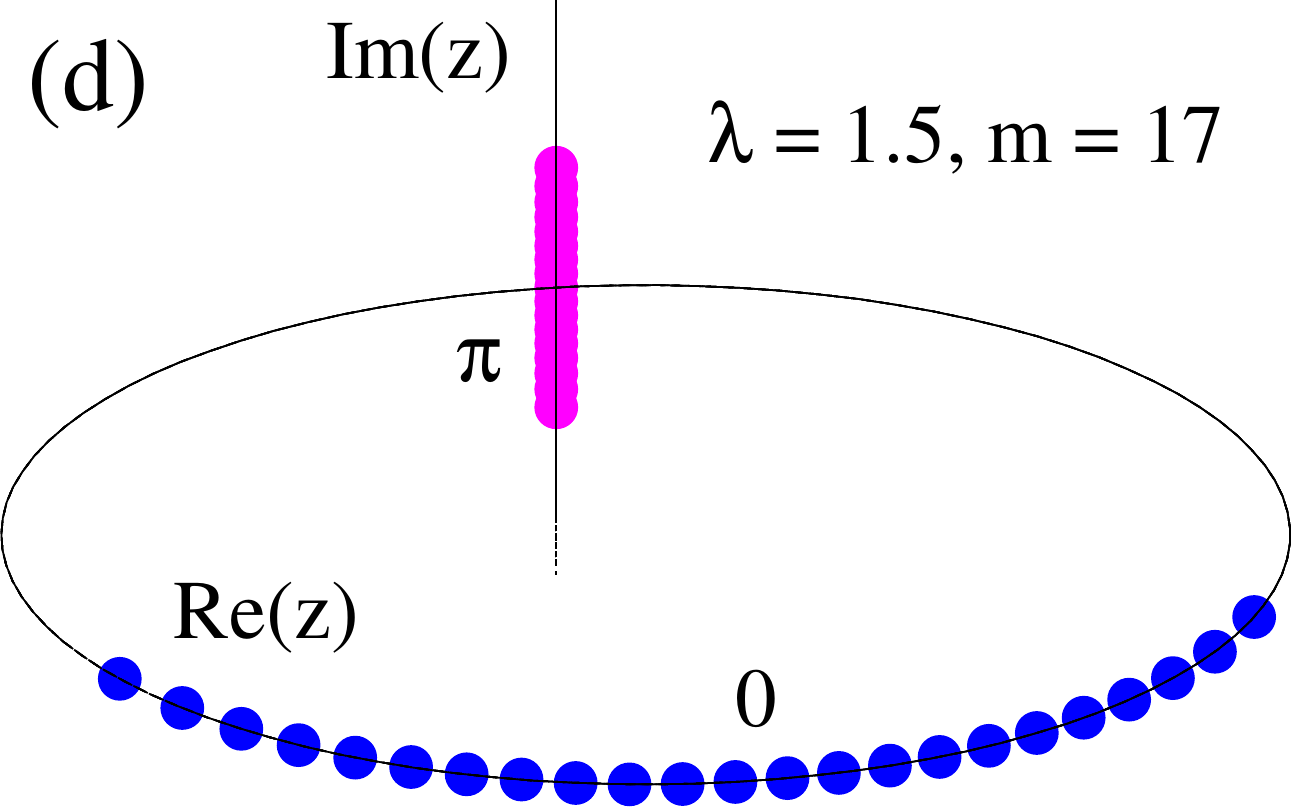}
\includegraphics[width=0.17\linewidth]{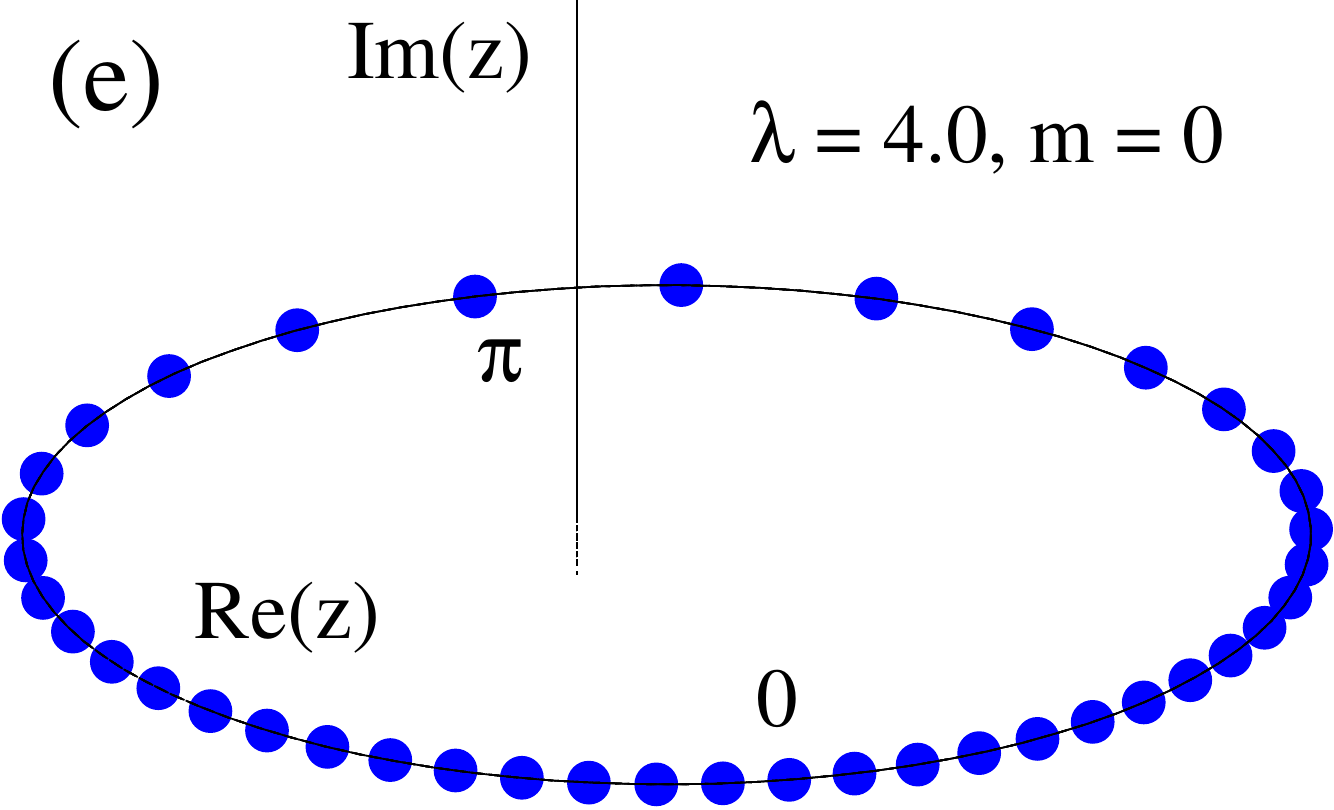}\\
\includegraphics[width=0.17\linewidth]{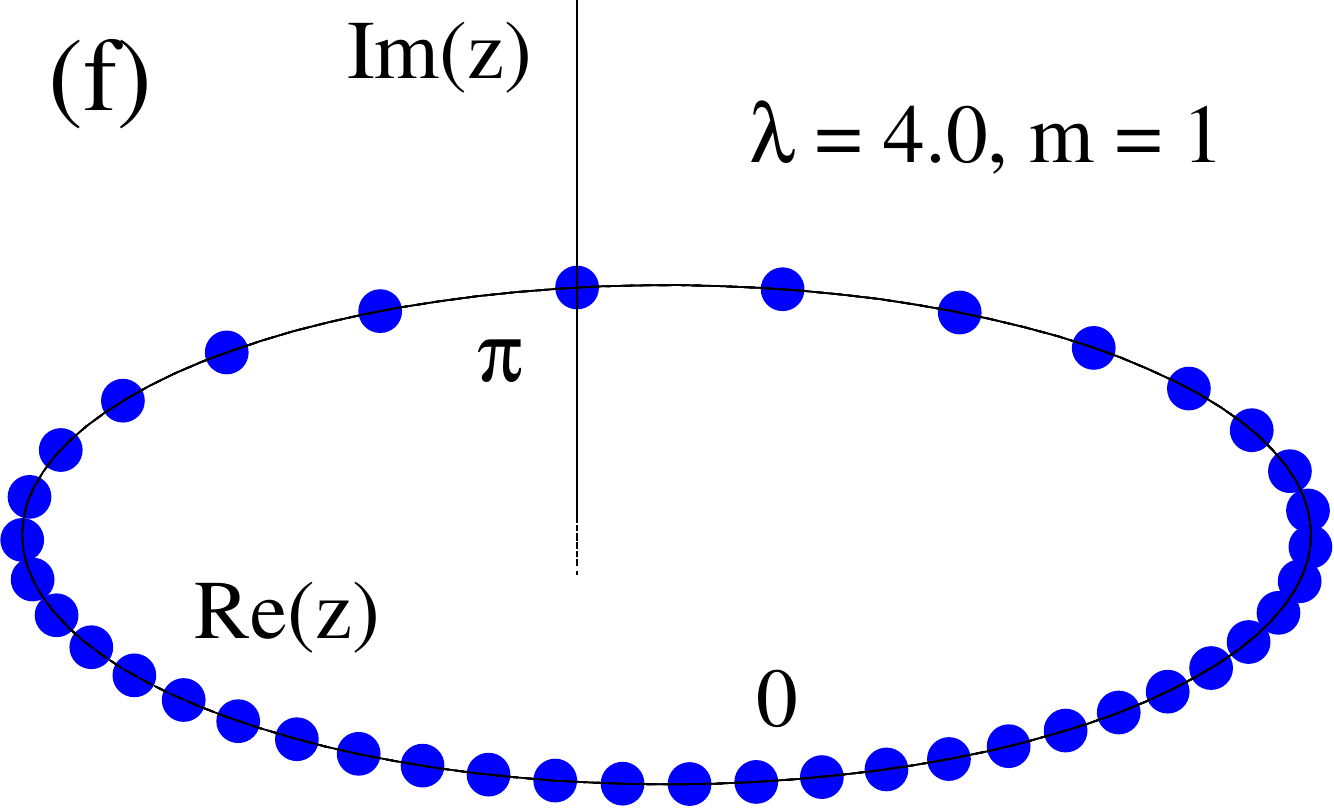}
\includegraphics[width=0.17\linewidth]{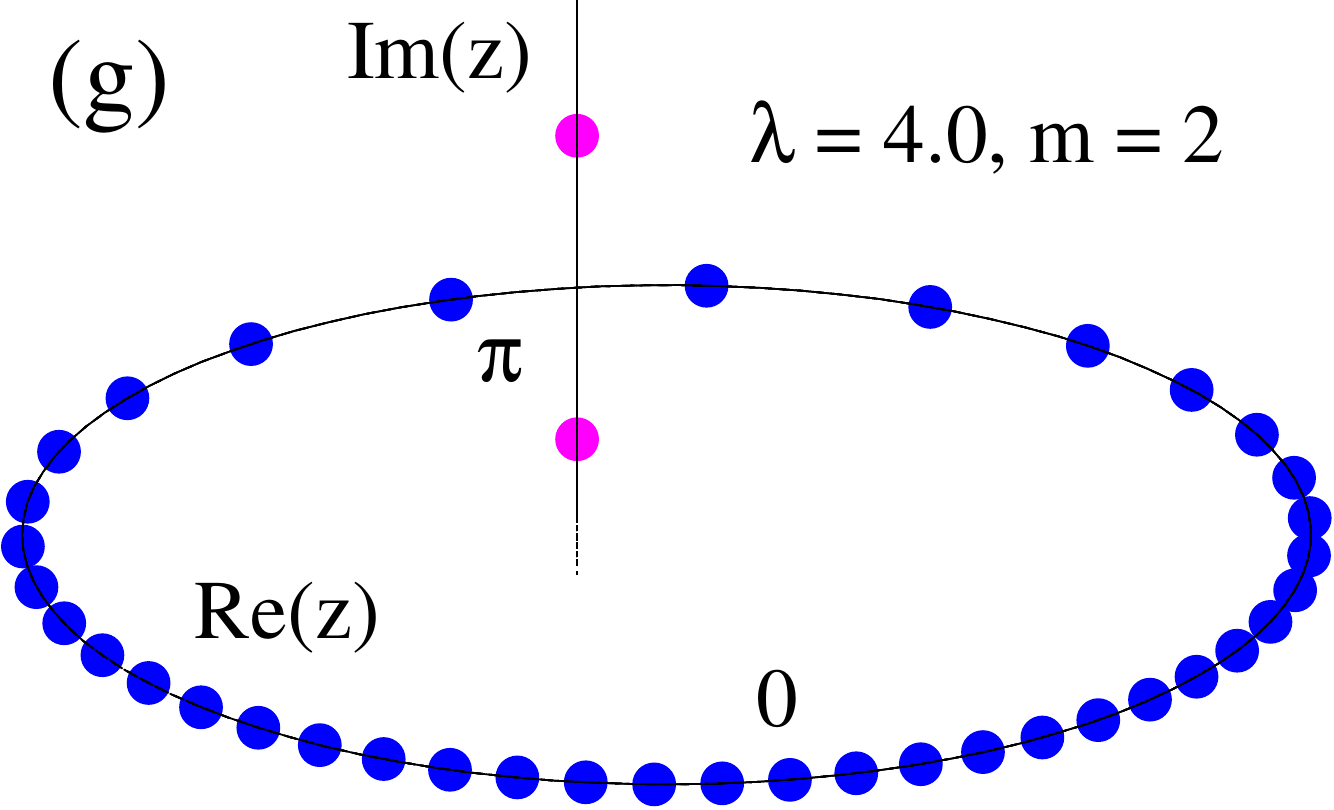}
\includegraphics[width=0.17\linewidth]{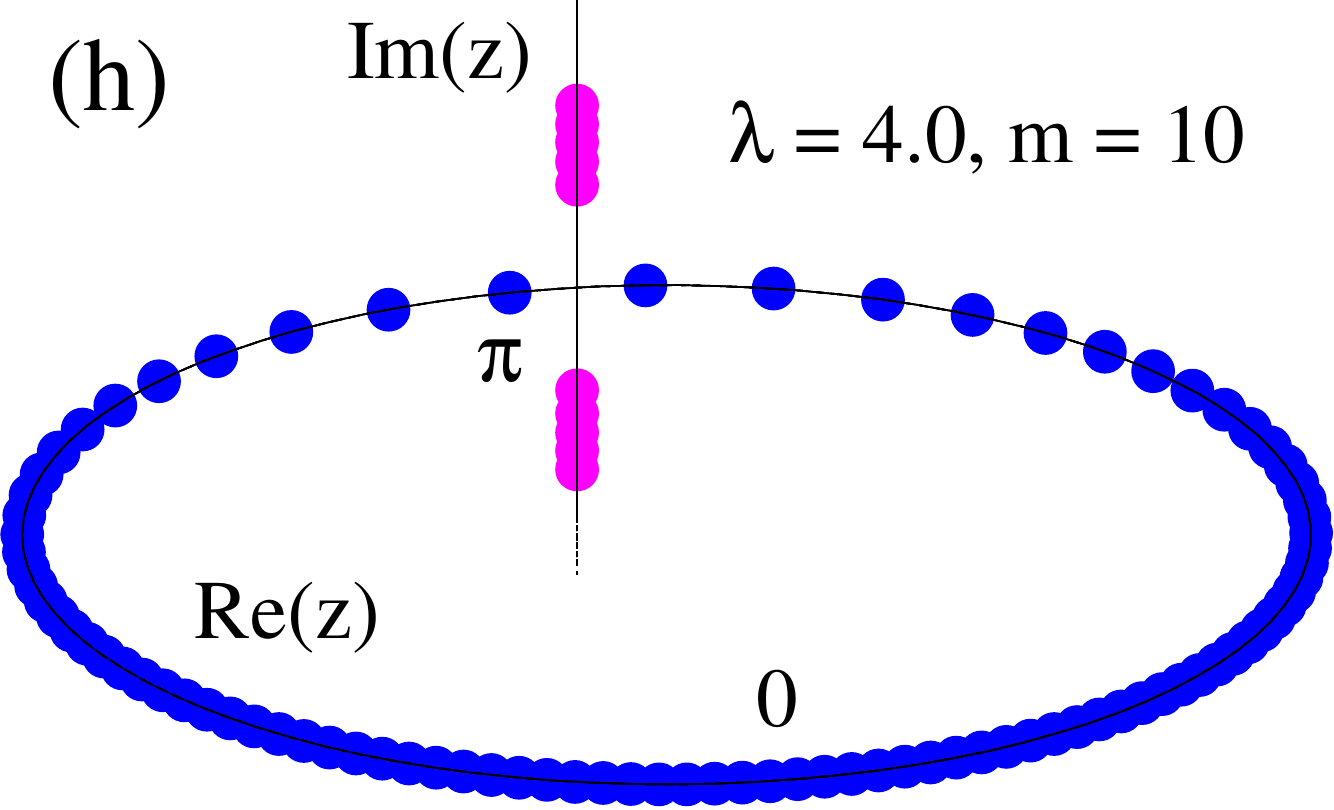}
\includegraphics[width=0.17\linewidth]{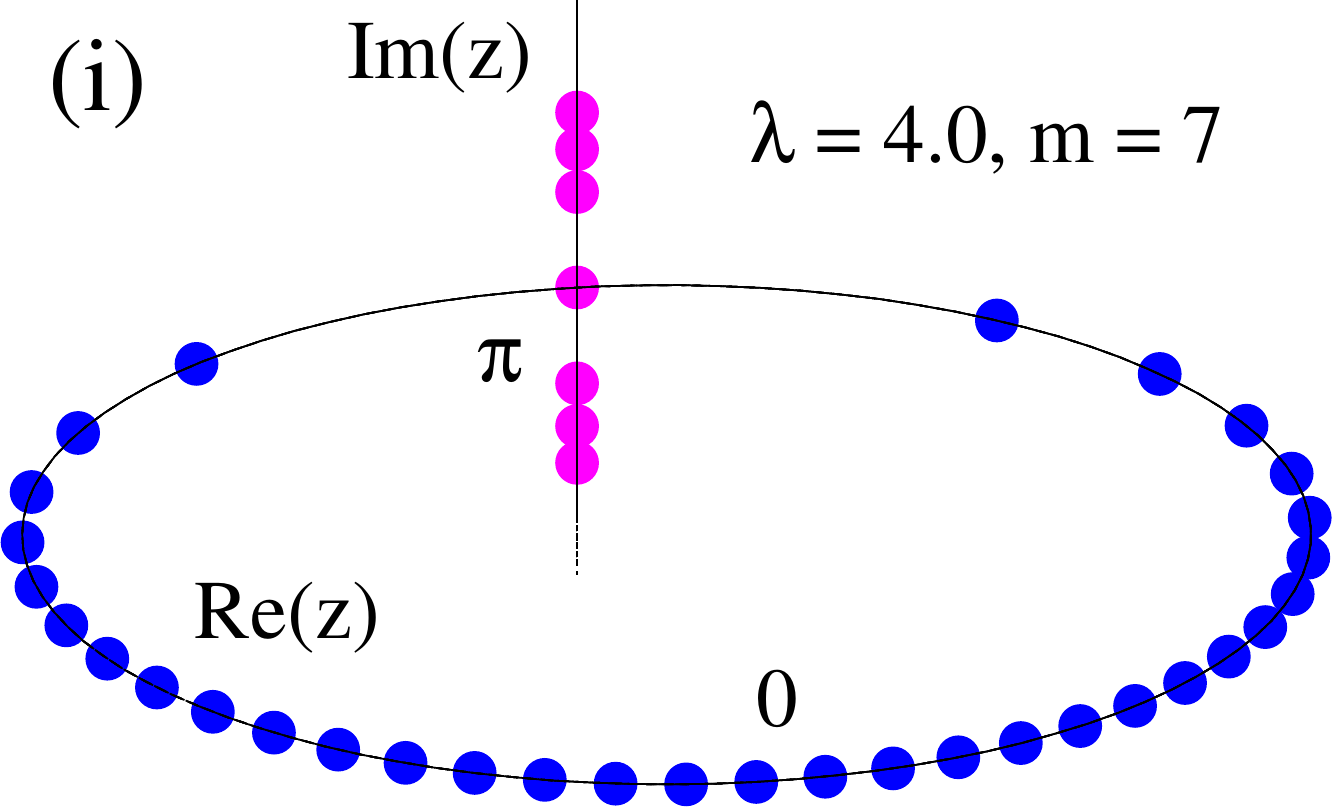}
\includegraphics[width=0.17\linewidth]{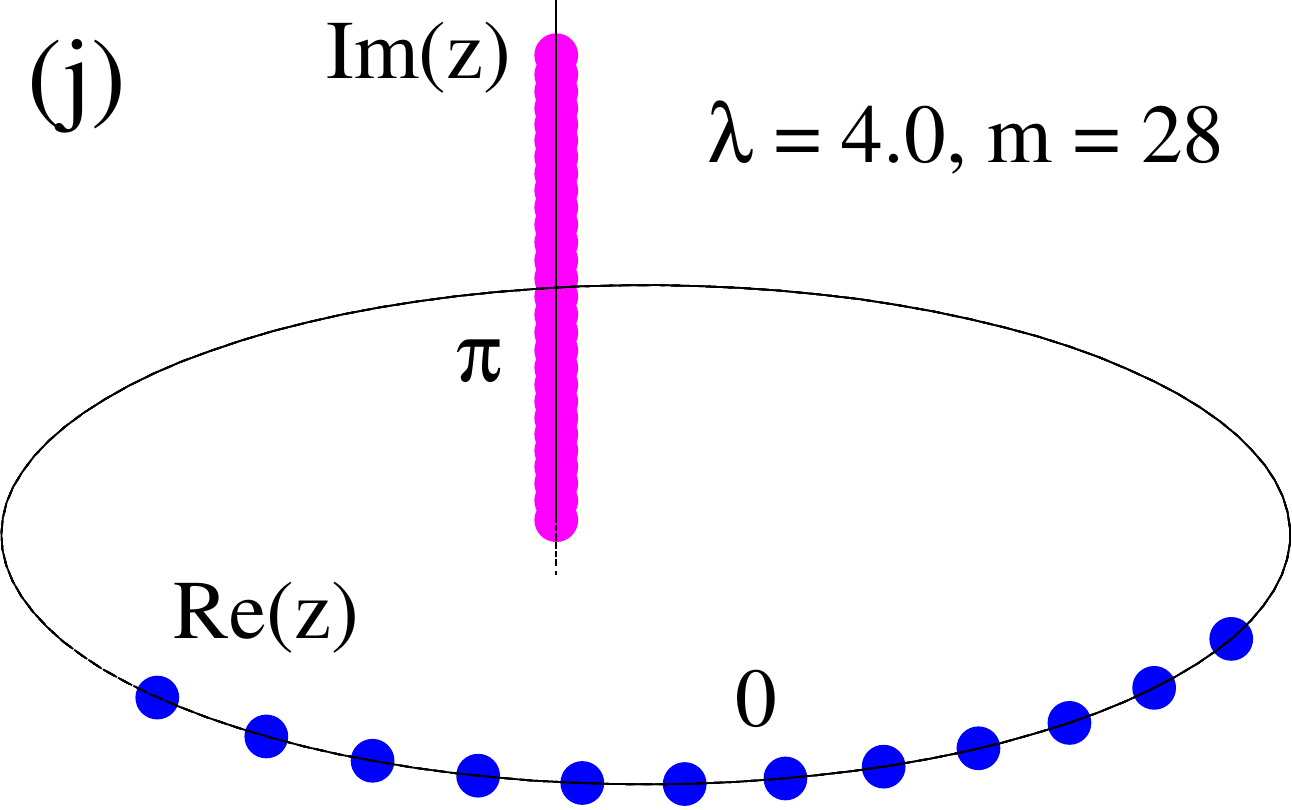}\\
  \caption{Saddle point configurations of eigenvalues $z_i$ in the weak-coupling (plots (a) - (d)), and in the strong-coupling (plots (e) - (j)) phases with different ``instanton numbers'' $m$. $N = 40$ on all of the plots except for the plot (h), where we take $N = 100$ in order to illustrate the three-cut solution at large $m$ and strong coupling.}
  \label{fig:saddle_points}
\end{figure*}

\noindent{\bf Gross-Witten-Wadia (GWW)  model:} The partition function is the integral, $\Z = \int\mathcal{D}U \, \exp\lr{\frac{N}{\lambda}\Tr\lr{U + U^{\dag}}}$, over $N \times N$ unitary matrices $U\in U\lr{N}$. $\Z $ can be expressed in terms of the eigenvalues $e^{i z_i}$ of $U$ \cite{Gross:80:1,Wadia:12:1}:
\begin{eqnarray}
\label{eq:Action}
 \Z = \prod\limits_{i=1}^{N} \int\limits_{-\pi}^{\pi} dz_i \, e^{-S\lr{z_i}}, \,
 S\lr{z_i} = \sum\limits_i V\lr{z_i} - \ln\Delta^2\lr{z_i} ,
 \nonumber \\
 V\lr{z} = -\frac{2 N}{\lambda} \cos\lr{z}, \,
 \Delta\lr{z_i} = \prod\limits_{i<j} \sin\lr{\frac{z_i-z_j}{2}} . \hspace*{0.5cm}
\end{eqnarray}
As $N\to \infty$, the leading contribution is from a distribution of eigenvalues $z_i$ along the line $\re z \in \mpipi$, $\im z = 0$, with a density function $\rho\lr{z}$, such that the number of eigenvalues in the interval $\lrs{z, z + d z}$ is  $dn = N \rho\lr{z} dz$. Writing the action $S$ in terms of $\rho\lr{z}$ identifies the large parameter $N^2$ in the exponent of the integrand, motivating a saddle point analysis. At $N=\infty$ this model has a third-order phase transition at $\lambda_c = 2$, where the third derivative of the free energy $E_0\lr{\lambda} = -\log{\Z}/N^2$ is discontinuous \cite{Gross:80:1,Wadia:12:1}.

The GWW model is more than a toy model: it exhibits the generic
phenomenon of phase transitions driven by gap closing in eigenvalue
distributions \cite{Majumdar:13:1,Marino:08:1}, which is also accompanied by condensation of Lee-Yang zeros in the complex coupling space \cite{Koelbig:81:1} and which is common to numerous physical systems such as 2D continuum gauge theory \cite{Douglas:93:2,Matytsin:94:1} and four-dimensional gauge theory at large $N$ \cite{Neuberger:06:1}, string theory \cite{Periwal:90:1,Gross:93:2}, large-N Chern-Simons theory \cite{Wadia:13:1}, general unitary and Hermitean matrix models \cite{Marino:08:1,Schiappa:15:1}, and
applications in mesoscopic conductance \cite{Vivo:08:1} and entanglement entropy \cite{Nadal:10:1}.

\noindent{\bf Complex Saddles in the GWW model:} We numerically solve the saddle equations, $\frac{\partial S}{\partial z_i} = 0$, $z_i \in \mathbb{C}$, at large but finite $N$. We use the next-order improved Newton iterations, the Halley method \cite{WeissteinHalleysMethod} \footnote{See Supplemental Material for a detailed description of this method}. We find complex saddles with novel properties not directly visible at $N = \infty$.
 \begin{figure}[h!tpb]
\centering
 \includegraphics[width=0.90\linewidth]{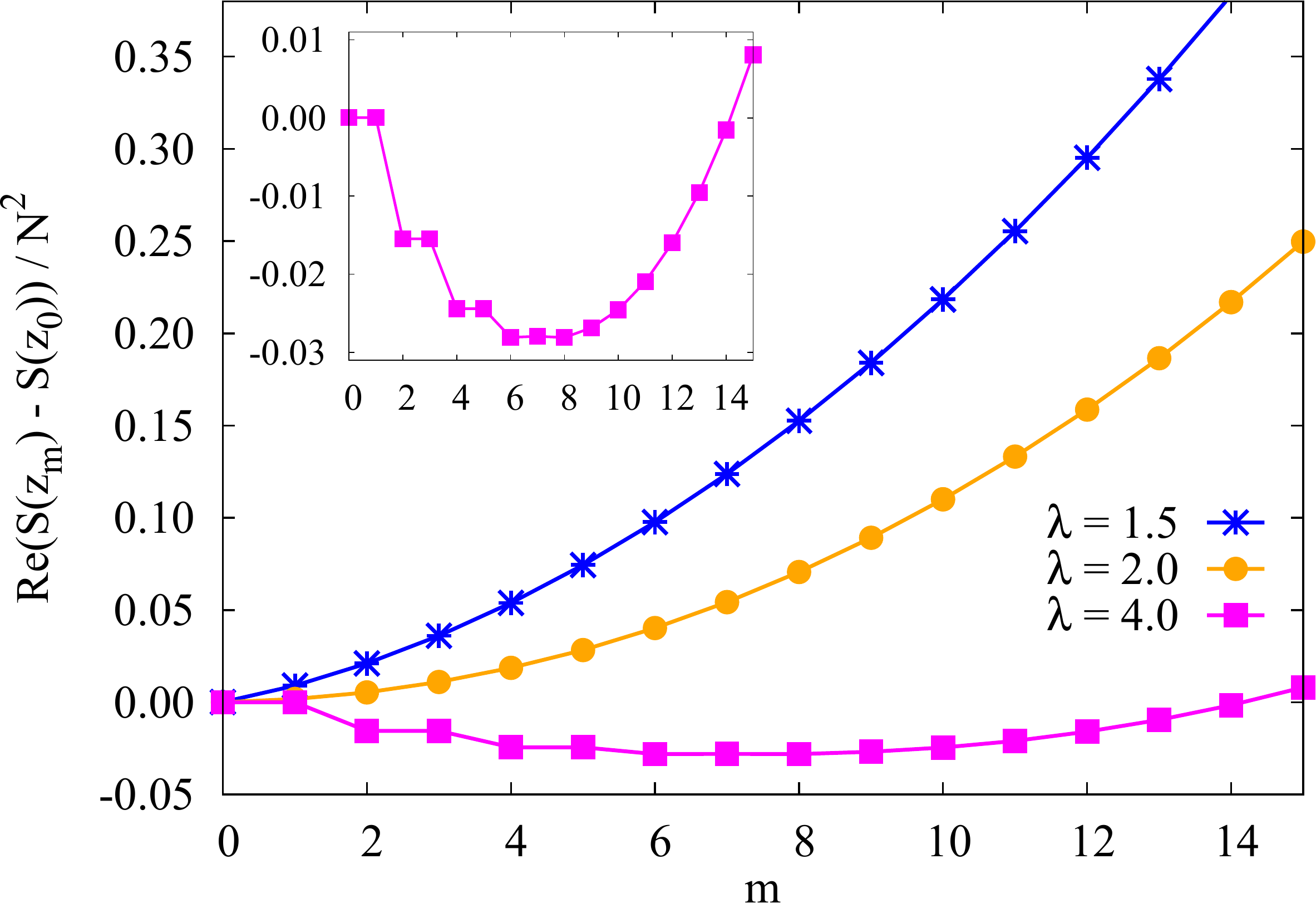}\\
 \caption{Real part of the saddle action, $\re S\lr{z}$, versus instanton number $m$, for different values of $\lambda$, at $N = 40$. The inset shows $\re S\lr{z}$ vs $m$ at $\lambda = 4$ on a larger scale.}
\label{fig:ReS_vs_m}
\end{figure}
\\ \emph{In both phases}, we find saddle configurations $z_i$ consisting of $(N - m)$ real eigenvalues located on the line $\re z \in \mpipi$, $\im z = 0$, and $m$ complex eigenvalues on the line $z = \pi + i y$, $y \in \mathbb{R}$. These lines are the steepest ascent contours of the potential $V\lr{z}$, originating from its extrema at $z = 0$ and $z = \pi$. The saddle configurations of $z_i$ are all symmetric with respect to these points; so for odd $m$ there is always one eigenvalue exactly at $z = \pi$.  Examples of saddle configurations in both phases are shown in Fig.~\ref{fig:saddle_points}, for various values of $m$. The action for these saddles has a real part plotted in Fig.~\ref{fig:ReS_vs_m} as a function of $m$, for three different values of $\lambda$: below, at, and above the phase transition. The imaginary part of $S\lr{z}$ is always a multiple of $\pi$: $\im S\lr{z} = \pi \floor{m/2}$, where $\floor{\cdot}$ is the floor function, so that the weight $\exp\lr{-S\lr{z}}$ is always real, but can have either sign. This sign comes exclusively from the Vandermonde determinant $\Delta^2\lr{z}$, and is interpreted as a hidden topological angle \cite{Unsal:15:3}.
  \begin{figure}[h!tpb]
\centering
 \includegraphics[width=0.90\linewidth]{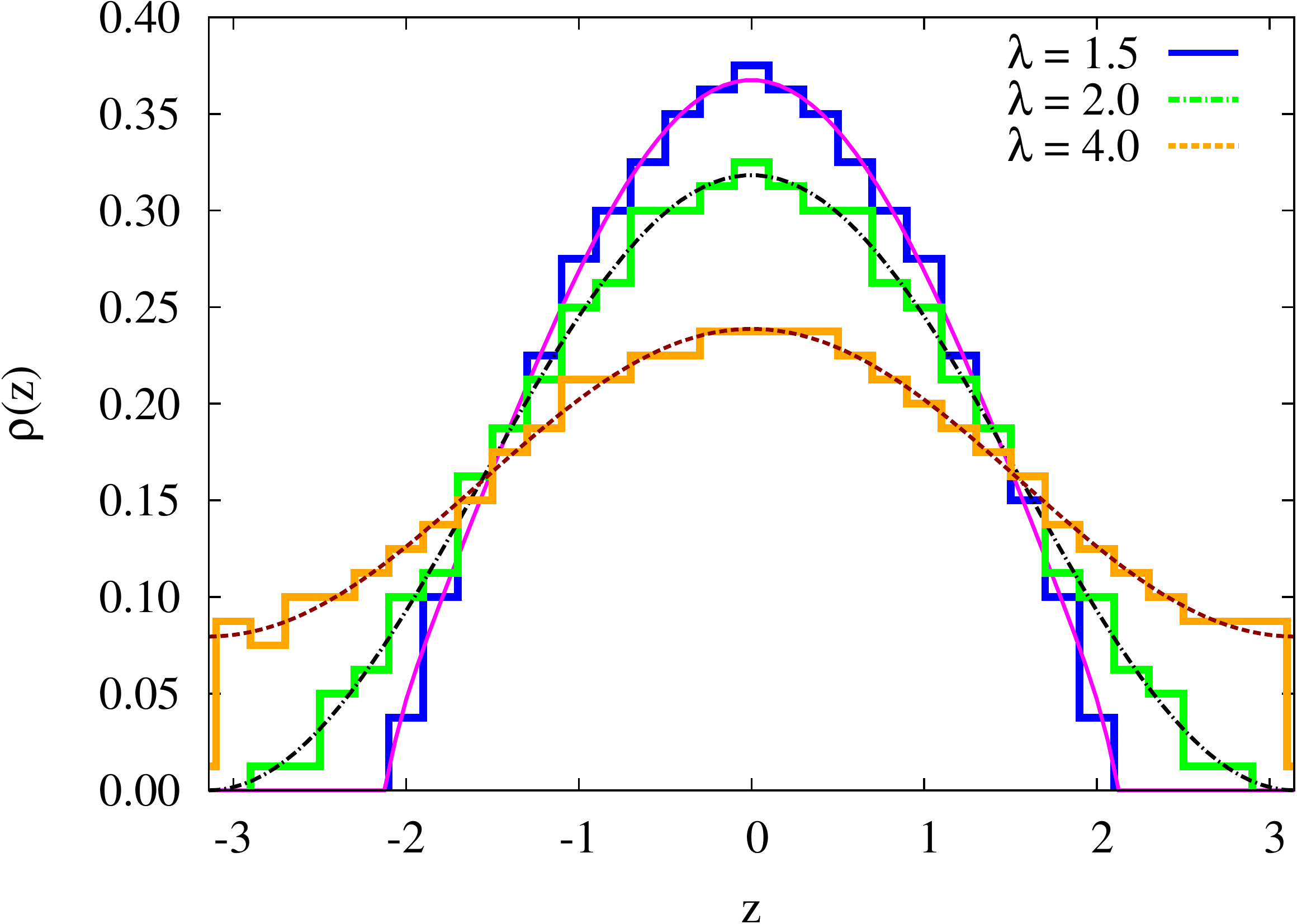}\\
 \caption{Numerical eigenvalue distributions for the $m=0$ saddle [$N = 400$], compared with the analytic  $\rho\lr{z}$ in (\ref{eq:weak-dist}, \ref{eq:strong-dist}).}
\label{fig:fig3}
\end{figure}

 \emph{Vacuum saddle:} we identify the $m=0$ saddle with the planar ($N=\infty$) contribution. As seen in Fig. \ref{fig:saddle_points}, in the weak-coupling phase, the $m=0$ saddle has a gapped distribution of real eigenvalues localized around the stable point $z=0$. At the phase transition $\lambda= 2$, this distribution closes at $z=\pi$, becoming ungapped in the strong-coupling phase. As shown in Fig.~\ref{fig:fig3}, the numerical distribution of eigenvalues fits the $N=\infty$ forms \cite{Gross:80:1,Wadia:12:1},
\begin{eqnarray}
\rho^{(w)}\lr{z}&=&\frac{2}{\lambda \pi} \cos\lr{\frac{z}{2}} \sqrt{\frac{\lambda}{2} - \sin^2\lr{\frac{z}{2}}}
 ,\quad  \lambda<2
 \label{eq:weak-dist}
 \\
\rho^{(s)}\lr{z} &=& \frac{1}{2\pi} \lr{1+\frac{2}{\lambda} \cos\lr{z}} ,\quad \lambda > 2
\label{eq:strong-dist}
\end{eqnarray}
Thus, the numerical $m=0$ free energy, $-S_0/N^2$, shows the expected 3rd-order phase transition at $\lambda=2$.

\begin{figure}[h!tpb]
\centering
 \includegraphics[width=0.90\linewidth]{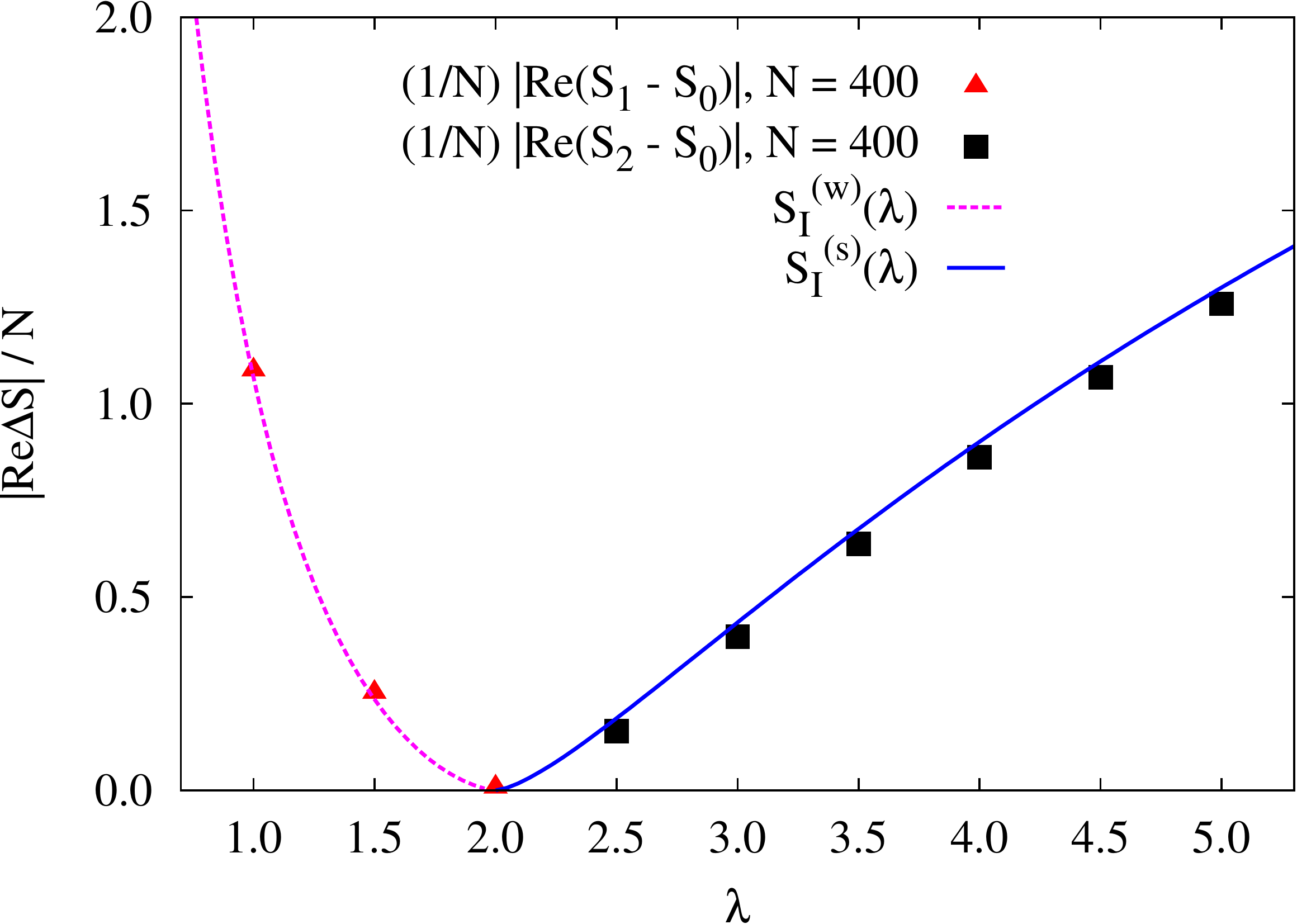}\\
 \caption{Numerical results (dots) for the relative actions of the leading non-vacuum saddles with $m=1$ (in the weak-coupling phase) and $m=2$ (in the strong-coupling phase), at $N = 400$. Solid lines are the analytic expressions (\ref{instanton_action_wc}) and (\ref{instanton_action_sc}).}
\label{fig:fig4}
\end{figure}

\begin{figure}[h!tb]
\centering
  \includegraphics[width=0.98\linewidth]{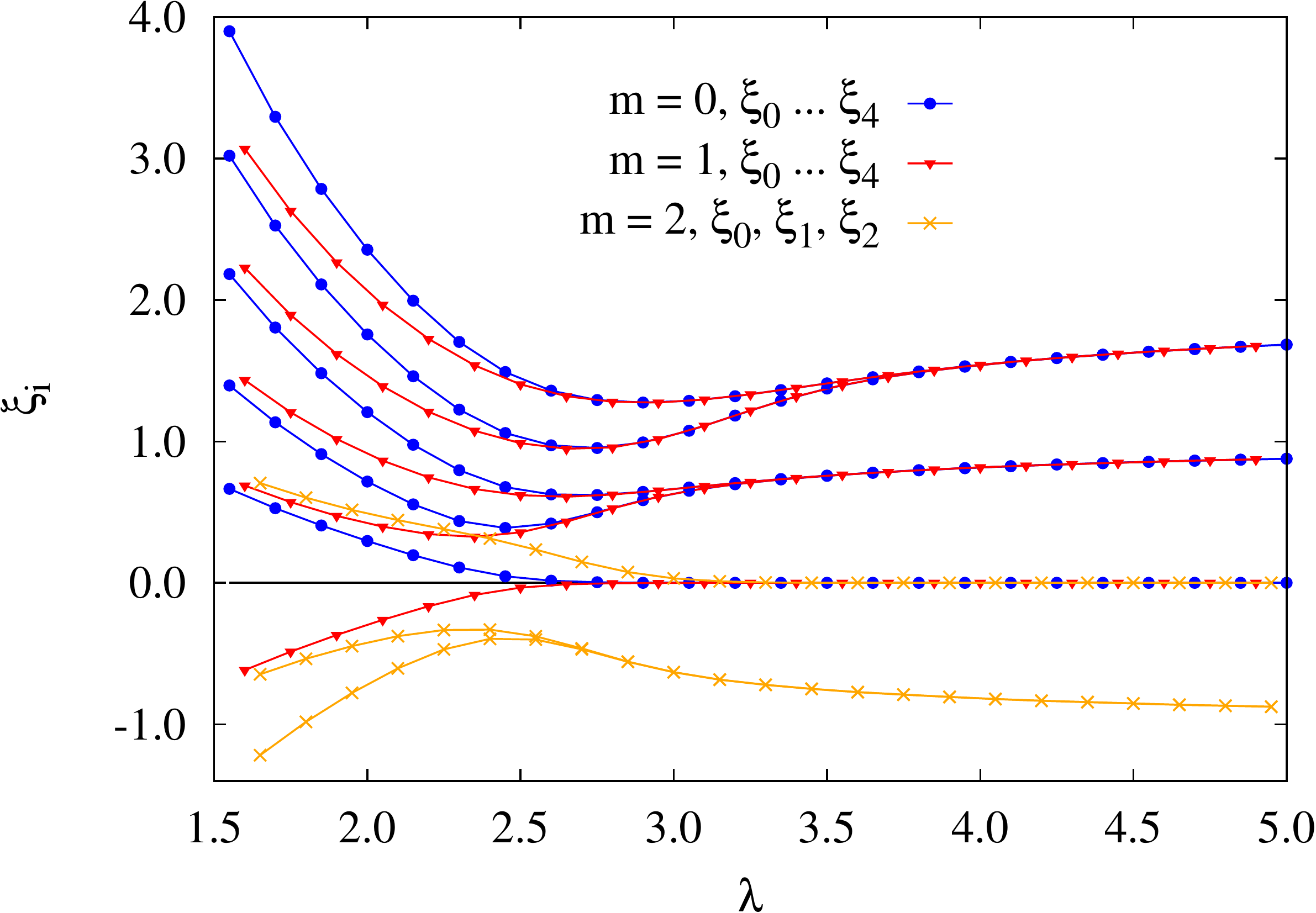}
  \caption{Several lowest eigenvalues $\xi_i$ of the Hessian matrix $H_{m, \,ij} = \frac{\partial^2 S_m}{\partial z_i \, \partial z_j}$ for saddles with $m = 0, 1, 2$ at $N = 40$. There are $m$ negative modes, and at strong coupling all modes, except the quasi-zero-mode, become quasi-degenerate.}
  \label{fig:hessian_eigenvalues}
\end{figure}

  \emph{Non-Vacuum Saddles at Weak-Coupling:} For $\lambda<2$, the lowest action non-perturbative saddle has $m=1$, with one eigenvalue at $z=\pi$, and has real action (relative to the vacuum action) exactly matching the weak-coupling instanton action $S_I^{(w)}$ \cite{Marino:08:1} (see Fig.~\ref{fig:fig4})
 \begin{eqnarray}
\label{instanton_action_wc}
 S_I^{(w)} = 4/\lambda \sqrt{1 - \lambda/2} - \arccosh\lr{(4 - \lambda)/\lambda}, \quad \lambda < 2 .
\end{eqnarray}
As $m$ increases, $m$ eigenvalues  line up along the imaginary direction $z=\pi+i y$, forming a cut (see Fig. \ref{fig:saddle_points}). This is a numerical indication of ``eigenvalue tunneling'', but we note that the tunneled eigenvalues are complex. For small $m\ll N$,  we find the conventional picture of a dilute instanton gas, with the real part of the action lowest at $m = 0$, and scaling approximately linearly with $m$, as $\re\lr{S_m-S_0} = m \, N \, S_I^{(w)}$ for $m \ll N$, with $S_I^{(w)}$ the weak-coupling instanton action (\ref{instanton_action_wc}). We thus identify the integer $m$, the number of eigenvalues along the imaginary direction, with the instanton number in the weak coupling phase. The $m$-instanton saddles have Hessian fluctuation matrices, $H_{m, \, ij} = \frac{\partial^2 S_m}{\partial z_i \, \partial z_j}$, with $m$ negative modes (see Fig. \ref{fig:hessian_eigenvalues}). Thus, the $m=1$ saddle gives an imaginary contribution to the saddle expansion of the free energy; we have confirmed that this is canceled by an imaginary term from the Borel summation of the divergent fluctuations about the $m=0$ vacuum saddle, a clear indication of resurgent cancellations. This can be traced to the resurgent asymptotics of individual Bessel functions, using the determinant representation \cite{Gross:80:1,Wadia:12:1} of the partition function: ${\mathcal Z}={\rm det}\left(I_{j-k}(2N/\lambda)\right)$ .

 \emph{Saddle Condensation Phase Transition:} As $\lambda \to 2$ from the weak-coupling side, the gap in the real part of the eigenvalue distribution closes at the unstable point ($z=\pi$) (see Figs. \ref{fig:saddle_points}, \ref{fig:fig3}). Furthermore, as seen in  Fig.~\ref{fig:ReS_vs_m}, the real part of the saddle action, relative to the vacuum value, tends to zero, so that all instantons with $m \ll N$ become equally important at the transition point, signaling instanton condensation \cite{Neuberger:81:1,Matytsin:94:1,Marino:08:1}.

\begin{figure}[h!tb]
\centering
  \includegraphics[width=0.98\linewidth]{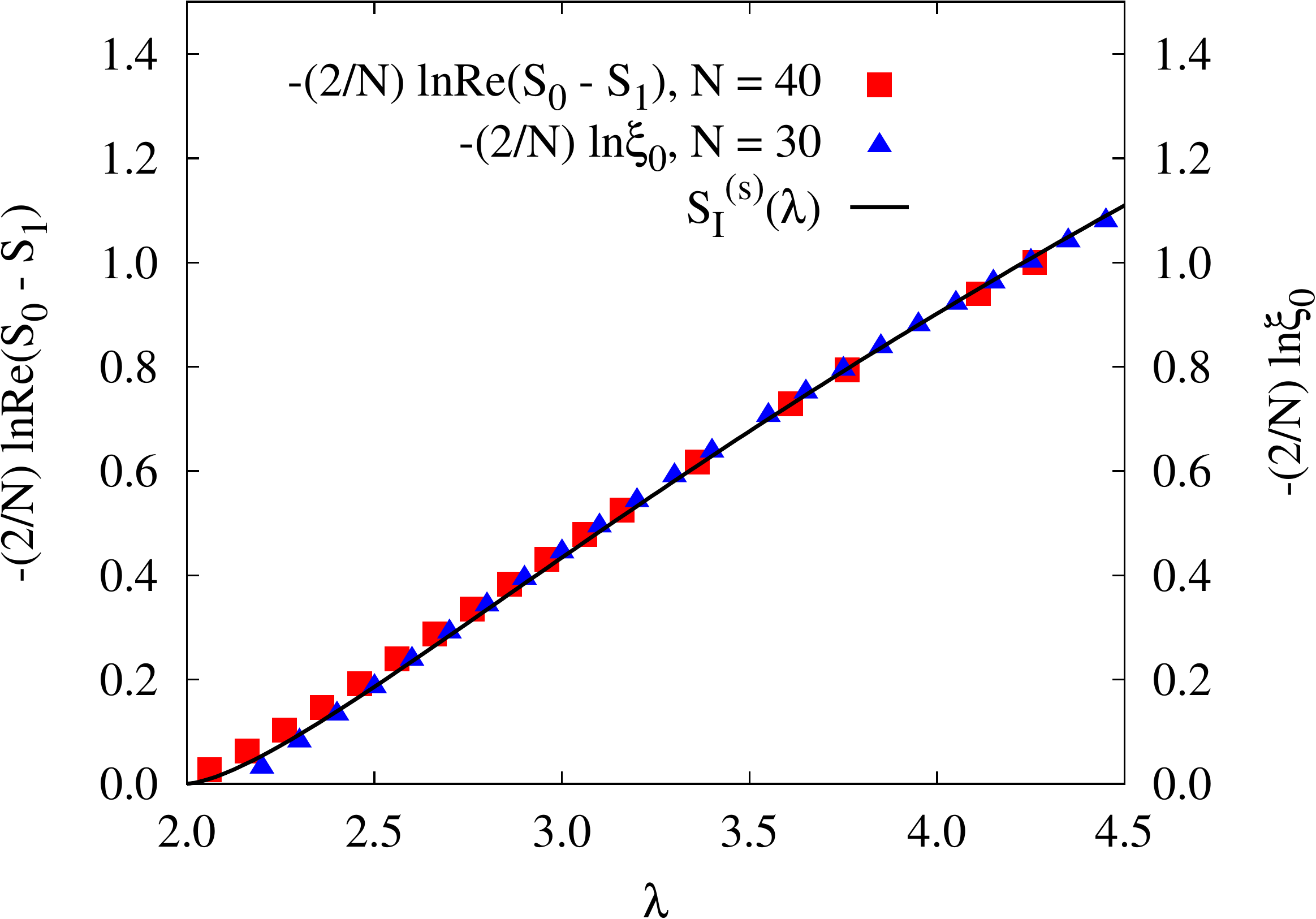}
  \caption{Comparison of the log of the difference between the actions of the $m=0$ and $m=1$ saddles (left vertical axis), and the log of the lowest Hessian eigenvalue $\xi_0$ (right vertical axis), in the strong coupling phase, with half the strong coupling instanton action in (\ref{instanton_action_sc}). Both effects are governed by the same exponential.}
  \label{fig:exp_splitting}
\end{figure}

\emph{Non-Vacuum Saddles at Strong-Coupling:} Since the unstable point $z = \pi$ is already in the support of $\rho^{(s)}(z)$, in the conventional picture non-vacuum saddles can no longer be constructed by dragging eigenvalues to $z = \pi$. Nevertheless, Mari\~no obtained the following strong-coupling ``instanton action" using a  trans-series ansatz in the string equation \cite{Marino:08:1} (see also Appendix B in \cite{Goldschmidt:80:1}):
\begin{eqnarray}
\label{instanton_action_sc}
 S_I ^{(s)}= 2 \arccosh\lr{\lambda/2} - 2 \sqrt{1 - 4/\lambda^2}, \quad \lambda \geq 2 .
\end{eqnarray}
Our numerical approach yields a natural interpretation of this ``instanton'' as a saddle configuration, with complex eigenvalue tunneling from the real to the imaginary axis (see Fig.~\ref{fig:saddle_points}). As in the weak coupling phase, $m$ eigenvalues line up along the imaginary direction, but these strong-coupling saddles have some surprising properties:

 (i) In the strong-coupling phase, at large $N$, the $m=1$ saddle has real action degenerate with that of the $m=0$ saddle, up to exponentially small corrections precisely of the form $\expa{-N/2 \, S_I^{(s)}}$, where $S_I^{(s)}(\lambda)$ is the strong-coupling instanton action (\ref{instanton_action_sc}), see Fig. \ref{fig:exp_splitting}. Physically, this is due to a quasi-zero mode in the strong-coupling regime. The $m = 0$ and $m = 1$ configurations have the same continuous eigenvalue density, but microscopically  differ by the presence or absence of a single eigenvalue at $z = \pi$ (see Fig.~\ref{fig:saddle_points}, (e) and (f)). To leading order in $1/N$, they can be related by a shift of every eigenvalue to the middle of the interval to its neighboring eigenvalue. At large $N$ this interval is inversely proportional to the density function $\rho\lr{z}$, so the shift of all eigenvalues by $\delta z_i \sim 1/\rho\lr{z_i}$ is a flat direction of the action. Correspondingly, at $N \rightarrow \infty$, $\delta z_i$ is the eigenvector of the Hessian $H_{ij} = \frac{\partial^2 S}{\partial z_i \, \partial z_j}$ with zero eigenvalue \footnote{See Supplemental Material for an explicit proof.}. Numerically we have found that as $N \rightarrow \infty$, the lowest eigenvalue $\xi_0$ vanishes exponentially fast as $\expa{-N/2 \, S_I^{(s)}}$ (see Fig.~\ref{fig:hessian_eigenvalues} and Fig.~\ref{fig:exp_splitting}). Interestingly, this is the same exponential factor seen in the splitting $\re\lr{S_1-S_0}$.

(ii) At strong-coupling, it is not the $m=1$ saddle, but rather the  $m=2$ saddle which we identify as the ``strong-coupling instanton'' configuration. This saddle is manifestly complex (Fig.~\ref{fig:saddle_points} (g)). It has action with real part equal to, as a function of $\lambda$,  the modulus of the strong-coupling action (\ref{instanton_action_sc}): $|\re\lr{S_2-S_0}| =  N S_I^{(s)}(\lambda)$, as shown in Fig. \ref{fig:fig4}. This reversal of sign is a numerical example of a phenomenon found in the context of the Painlev\'e equations, where formal trans-series arise with saddles of both signs of the action \cite{Marino:12:1,Schiappa:15:3,Schiappa:12:1}.

(iii) At strong-coupling, as $m$ increases more eigenvalues move away from the real axis, forming a distinct two-cut structure around $z = \pi$  (with one eigenvalue in the gap at $z = \pi$ if $m$ is odd): see Fig.~\ref{fig:saddle_points} (h). The real part of the action decreases with $m$ until it reaches a critical value $m^\star$, after which it increases again: see  Fig.~\ref{fig:ReS_vs_m}. When $m$ reaches  $m^\star$ the gap between the two cuts closes, and at the same time the distribution of the remaining eigenvalues on the real axis becomes gapped (Fig.~\ref{fig:saddle_points} (i), (j)). The saddle-point action scales linearly with $m$ for $m \ll m^{\star}$: $|\re\lr{S_m-S_0}| \approx \floor{m/2} \, N \, S_I^{(s)}$, where $S_I^{(s)}$ is the strong-coupling instanton action (\ref{instanton_action_sc}). Note the floor function in this expression, which implies that the aforementioned degeneracy of the action for $m=0$ and $m=1$ persists also for the pairs of saddle points with $m = 2 n$ and $m = 2n + 1$, for $m<m^\star$ (see the ``stairs'' at low $m$ in the inset in Fig.~\ref{fig:ReS_vs_m}). Correspondingly, the Hessian matrices for all saddles with $m < m^{\star}$ have quasi-zero-modes, vanishing exponentially with $N$ (see Fig. \ref{fig:hessian_eigenvalues}), as for the $m=1$ saddle, but with an $m$-dependent pre-factor.

(iv) As in the weak coupling phase, at strong-coupling the Hessian matrix for the $m$-saddle has $m$ negative modes (see Fig. \ref{fig:hessian_eigenvalues}). But in the strong-coupling phase, all eigenvalues except the zero-mode become doubly degenerate, with degeneracy splitting governed again by the exponentially small quantity $\expa{-N/2 \, S_I^{(s)}(\lambda)}$.

Our numerical results indicate that the GWW partition function and free energy have trans-series expansions also in the strong-coupling phase, due to complex saddle points. This provides a (complex) saddle interpretation of Mari\~no's trans-series result from the string equation \cite{Marino:08:1}, and is also consistent with the double-scaling limit described by the McLeod-Hastings solution to the Painlev\'e II equation, valid near the phase transition. On the weak coupling side this solution has exponential corrections $\sim \exp(-N S_I^{(w)})$, while on the strong-coupling side the leading behavior is already exponential $\exp(-N/2 S_I^{(s)})$, which implies $\exp(-N S_I^{(s)})$ behavior for the free energy \cite{Marino:08:1}. Furthermore, deep in the strong-coupling region, with $\lambda\gg 2$, and using the method of orthogonal polynomials, Goldschmidt found \cite{Goldschmidt:80:1} corrections behaving like $\frac{1}{N^2} \left(\frac{\lambda}{e}\right)^{-2N}\sim \frac{1}{N^2}
\exp(- N S_I^{(s)}(\lambda))$ [note that
$S_I^{(s)}\sim 2\ln\left(\frac{\lambda}{e}\right)+\frac{2}{\lambda^2}+\dots$, for $\lambda\gg 2$].

\mysection{Conclusions}{sec:conclusions}
Our numerical study reveals a surprisingly rich structure of
complex-valued saddles in both the weak- and strong-coupling phases of two-dimensional
lattice gauge theory, represented by the Gross-Witten-Wadia unitary matrix
model. These complex saddles are intimately related to the resurgent structure of the 1/N expansion. We find a new complex saddle interpretation of Mari\~no's
strong-coupling instanton action, and these saddles have novel physical
properties. There is clear numerical evidence for instanton condensation
at the transition. In both phases, eigenvalue tunneling produces complex
saddles, and these results suggest a Lefschetz thimble interpretation of
the saddle point expansion. Given the direct relation between the instanton actions in the matrix model (\ref{eq:Action}) and in 2D continuum gauge theory \cite{Marino:08:1}, we expect similar results for complex-valued saddles to apply also to continuum 2D gauge theories \cite{Douglas:93:2,Matytsin:94:1}.
\begin{acknowledgments}
 We thank T.~Sulejmanpasic and M.~\"Unsal for interesting and stimulating discussions. This work was supported by the S.~Kowalevskaja award from Alexander von Humboldt Foundation (PB, SV), and the U.S. DOE grant DE-SC0010339 (GD).
\end{acknowledgments}

\clearpage
\appendix

\setcounter{equation}{0}
\setcounter{figure}{0}
\setcounter{table}{0}
\setcounter{page}{1}
\makeatletter
\renewcommand{\theequation}{S\arabic{equation}}
\renewcommand{\thefigure}{S\arabic{figure}}

\begin{widetext}

\begin{center}
\textbf{\large Supplemental Material}
\end{center}

\section{Numerical solution of the complexified saddle-point equations}
\label{apdx:Halley}

 Since the effective action $S\lr{z}$ of the Gross-Witten-Wadia matrix model is a holomorphic function of $N$ eigenvalues $z_i$ which can be in general complex, the saddle-point equations $\frac{\partial S\lr{z}}{\partial z_i} = 0$  can be solved numerically using e.g. Newton iterations, which are known to work well also for complex roots of holomorphic equations. However, in our case the simplest Newton iterations turned out to be unstable for values of $N$ larger than approximately $N = 20$. For this reason in this work we have used the next-order improvement of the Newton method, the so-called Halley method \cite{WeissteinHalleysMethod} in order to solve the saddle point equation:
\begin{eqnarray}
\label{eq:Halley_Iterations}
 z^{(n+1)}_i = z^{(n)}_i
 - 2 \lr{2 H\lrs{z^{(n)}_k} - J\lrs{z^{(n)}_k} H^{-1}\lrs{z^{(n)}_k} F\lrs{z^{(n)}_k}}^{-1} F\lrs{z^{(n)}_k}_i ,
\end{eqnarray}
Here we have defined
\begin{eqnarray}
\label{eq:Matrix_definitions}
 F_i\lrs{z_l}     = \frac{\partial S\lrs{z_l}}{\partial z_i}, \quad
 H_{ij}\lrs{z_l}  = \frac{\partial^2 S\lrs{z_l}}{\partial z_i \partial z_j}, \quad
 J_{ijk}\lrs{z_l} = \frac{\partial^3 S\lrs{z_l}}{\partial z_i \partial z_j \partial z_k} .
\end{eqnarray}
In practice, each iteration can be implemented as follows (we assume summation over repeated indices):
\begin{enumerate}
\label{eq:Halley_practical_implementation}
\item Solve $H_{ij}\lrs{z^{(n)}_k} u_j = F_i\lrs{z^{(n)}_k}$ for $u_i$,
\item Solve $\lr{H_{ij}\lrs{z^{(n)}_k} - \frac{1}{2} J_{ijk}\lrs{z^{(n)}_k} u_k} v_j = J_{ijk}\lrs{z^{(n)}_k} u_j u_k$ for $v_i$,
\item Finally, $z^{(n+1)}_i = z^{(n)}_i - u_i - v_i/2$.
\end{enumerate}
Note that it is not necessary to store the whole tensor $J_{ijk}$ in computer memory since we only need to know its contractions with the vectors $u$ and $v$. Using this method we were able to solve the holomorphic saddle-point equations for the values of $N$ as large as $N = 400$. As initial conditions for the Halley iterations we choose values $z_i^{\lr{0}}$ randomly distributed in the rectangle $\re{z_i} \in \mpipi$, $\im{z_i} \in \lrs{-5.0, 5.0}$, which yields sufficiently many distinct solutions with apriori unknown properties. We distinguish the solutions up to the obvious symmetry of arbitrary permutations of eigenvalues $z_i$.

\section{Eigenspectrum of the Hessian matrix}
\label{apdx:zero_mode}

In the main text, we have already mentioned that the Hessian matrices of the saddle points with $m < m^{\star}$ at $N \rightarrow \infty$ have $m$ negative eigenvalues, and one quasi-zero eigenvalue. Further, non-quasi-zero eigenvalues are all doubly quasi-degenerate. In this Supplementary Material we analytically prove the existence of the zero eigenvalue of the Hessian at $N = \infty$ and in the strong-coupling phase, and also consider the other eigenvalues of the Hessian matrix in more detail.

At $N \rightarrow \infty$ the equation on the eigenstate $\delta z\lr{z}$ of the Hessian matrix of the action \iftoggle{arxiv}{(\ref{eq:Action})}{of the Gross-Witten model (Eq.~(1) in the main text)} with zero eigenvalue can be written as the following linear integral equation:
\begin{eqnarray}
\label{eq:Eigenvector_equation}
 \lrs{H \delta z}\lr{z} =
 - \frac{2}{\lambda}\cos\lr{z}\delta z\lr{z}
 - P.V.\int\limits^\pi_{-\pi} \d z^\prime \rho\lr{z^\prime} \frac{\delta z\lr{z}-\delta z\lr{z^\prime}}{\sin^2\lr{\frac{z-z^\prime}{2}}} = 0 ,
\end{eqnarray}
where $P.V.$ stands for the principal value of the integral, and in the strong-coupling phase with $\lambda > 2$ the density function $\rho\lr{z}$ for the vacuum saddle is given by $\rho\lr{z} = \frac{1}{2\pi}\lr{1 + \frac{2}{\lambda}\cos\lr{z}}$. A direct substitution of $\delta z\lr{z} = 1/\rho\lr{z}$ into (\ref{eq:Eigenvector_equation}) solves this equation. Note that since $\delta z\lr{z}$ is analytic for $z \in \lrs{-\pi, \pi}$, the integrand in the second term on the right-hand side of (\ref{eq:Eigenvector_equation}) diverges only as $1/z$ rather than $1/z^2$, thus the integral is well defined in the sense of Cauchy principal value.

\begin{figure}[h!tb]
\centering
  \includegraphics[width=0.5\linewidth]{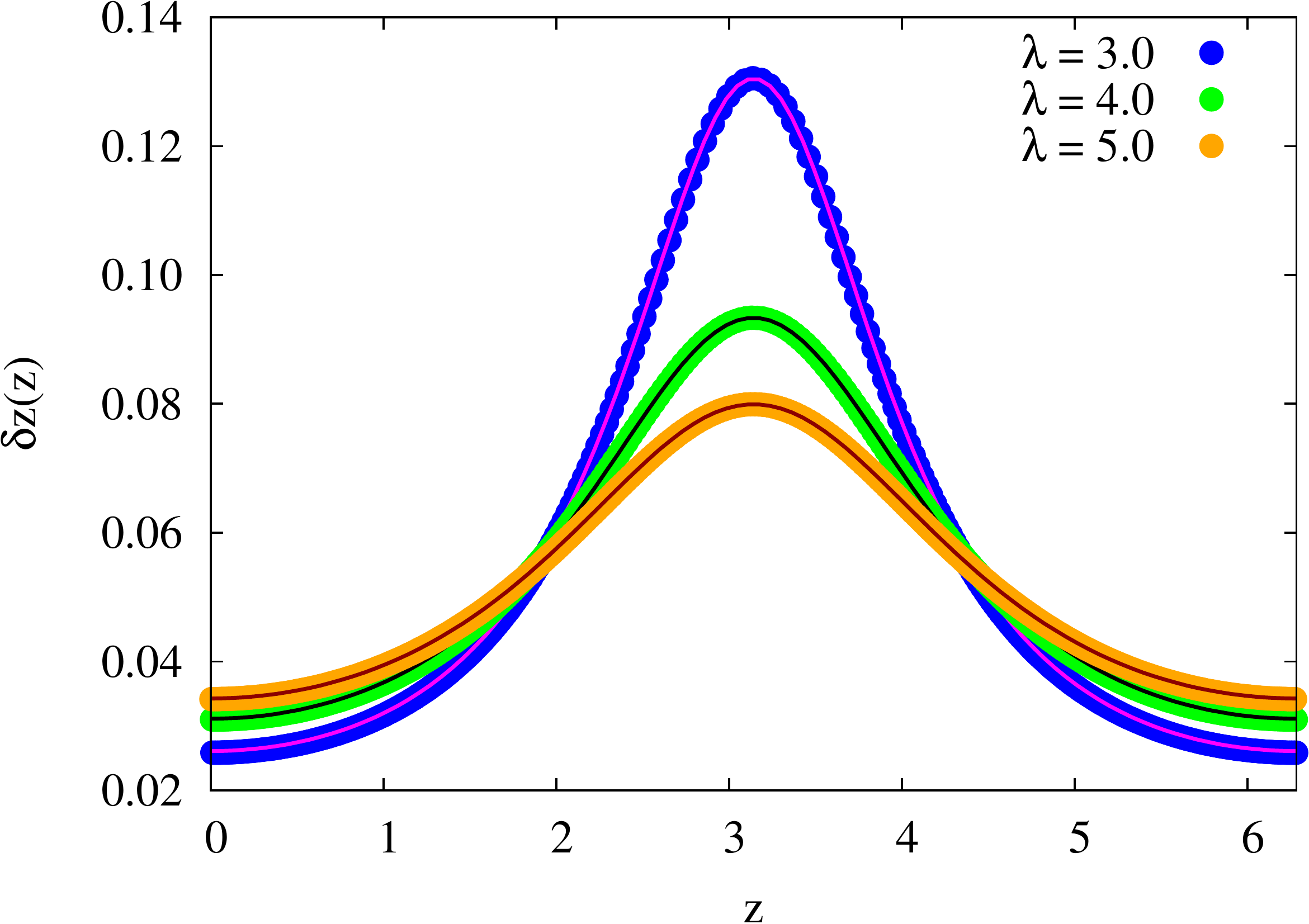}
  \caption{Comparison of the analytic expression $\delta z \sim 1/\rho^{(s)}\lr{z}$ for the zero mode of the Hessian matrix $H_0$ of the $m = 0$ saddle in the strong-coupling phase (solid line) with numerically calculated eigenvector of $H_0$ which corresponds to the lowest eigenvalue at $N = 400$.}
  \label{fig:Zero_mode}
\end{figure}

At finite $N$, we have found the eigensystem of the Hessian matrix $H_{ij}$ numerically using the \texttt{LAPACK} routine \texttt{zgeev}. First, on Fig.~\ref{fig:Zero_mode} we demonstrate a good agreement between the numerically calculated eigenvector of $H_{ij}$ with smallest (by absolute value) eigenvalue and the analytic expression $\delta z = 1/\rho^{(s)}(z)$, where $\rho^{(s)}\lr{z}\iftoggle{arxiv}{}{ = \frac{1}{2\pi} \lr{1+\frac{2}{\lambda} \cos\lr{z}}}$ is the strong-coupling eigenvalue distribution \iftoggle{arxiv}{(\ref{eq:strong-dist})}{}.

On the left plot on Fig.~\ref{fig:hessian_spectral_flow} we further demonstrate that at finite $N$ both the quasi-zero eigenvalue $\xi_0$ of $H$ and the splittings $\xi_1 - \xi_2$, $\xi_3 - \xi_4$, $\ldots$ between higher eigenvalues are all controlled by the same exponential $\expa{-N/2 \, S_I}$. Thus the lowest eigenvalue becomes exactly zero and the higher eigenvalues become exactly degenerate only at $N = \infty$. We also note that apart from the emerging degeneracy, the low-lying eigenvalues of the Hessian (or the degenerate pairs thereof in the strong-coupling phase) scale approximately as $N^1$, and the eigenvalues at the upper edge of the spectrum (see the right plot on Fig.~\ref{fig:hessian_spectral_flow}) - approximately as $N^2$. From the right plot on Fig.~\ref{fig:hessian_spectral_flow} we see also that the double degeneracy of eigenvalues is absent at the upper edge of the spectrum.

 \begin{figure*}[h!tpb]
  \centering
   \includegraphics[width=0.45\linewidth]{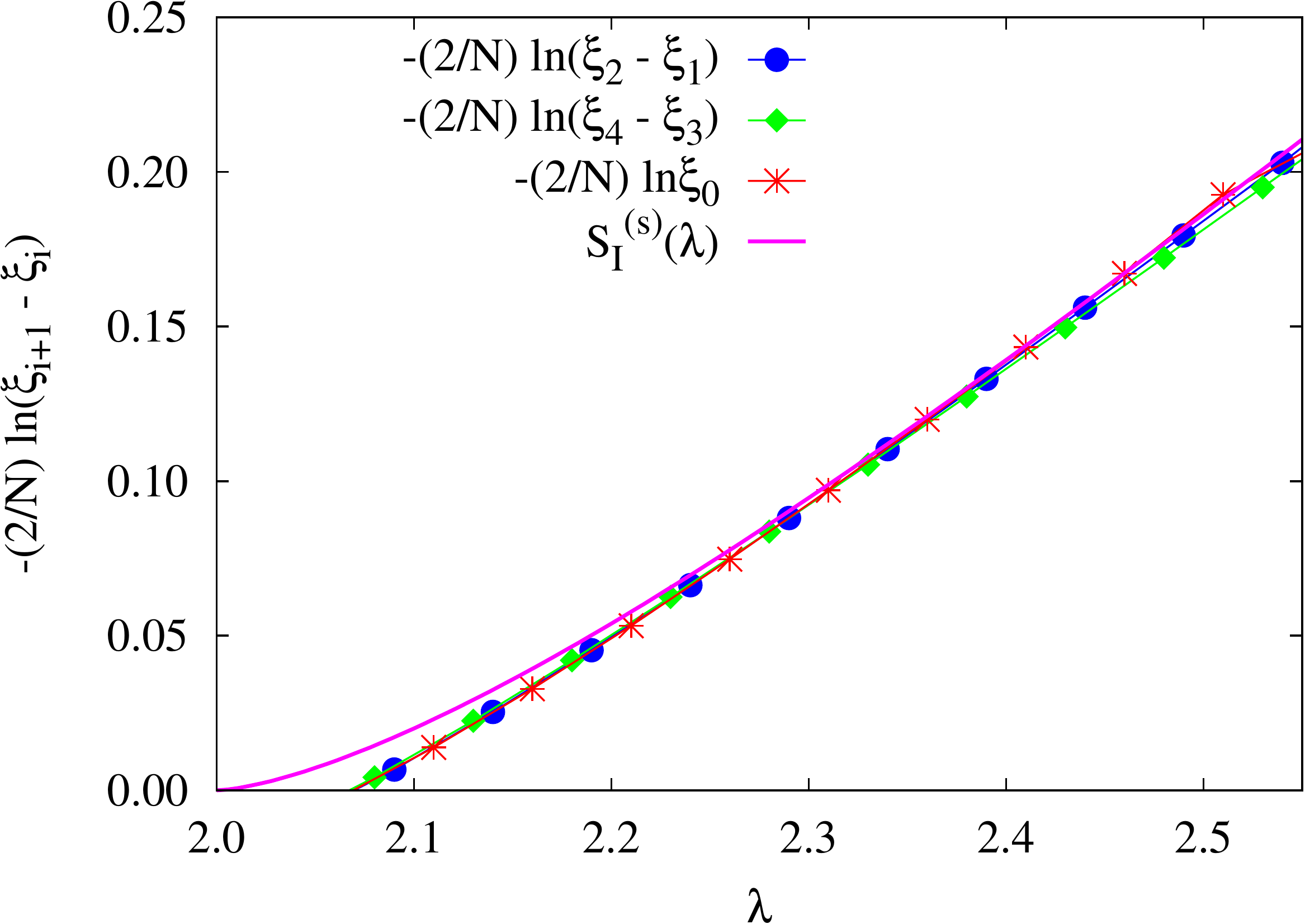}
   \hspace{0.05\linewidth}
   \includegraphics[width=0.45\linewidth]{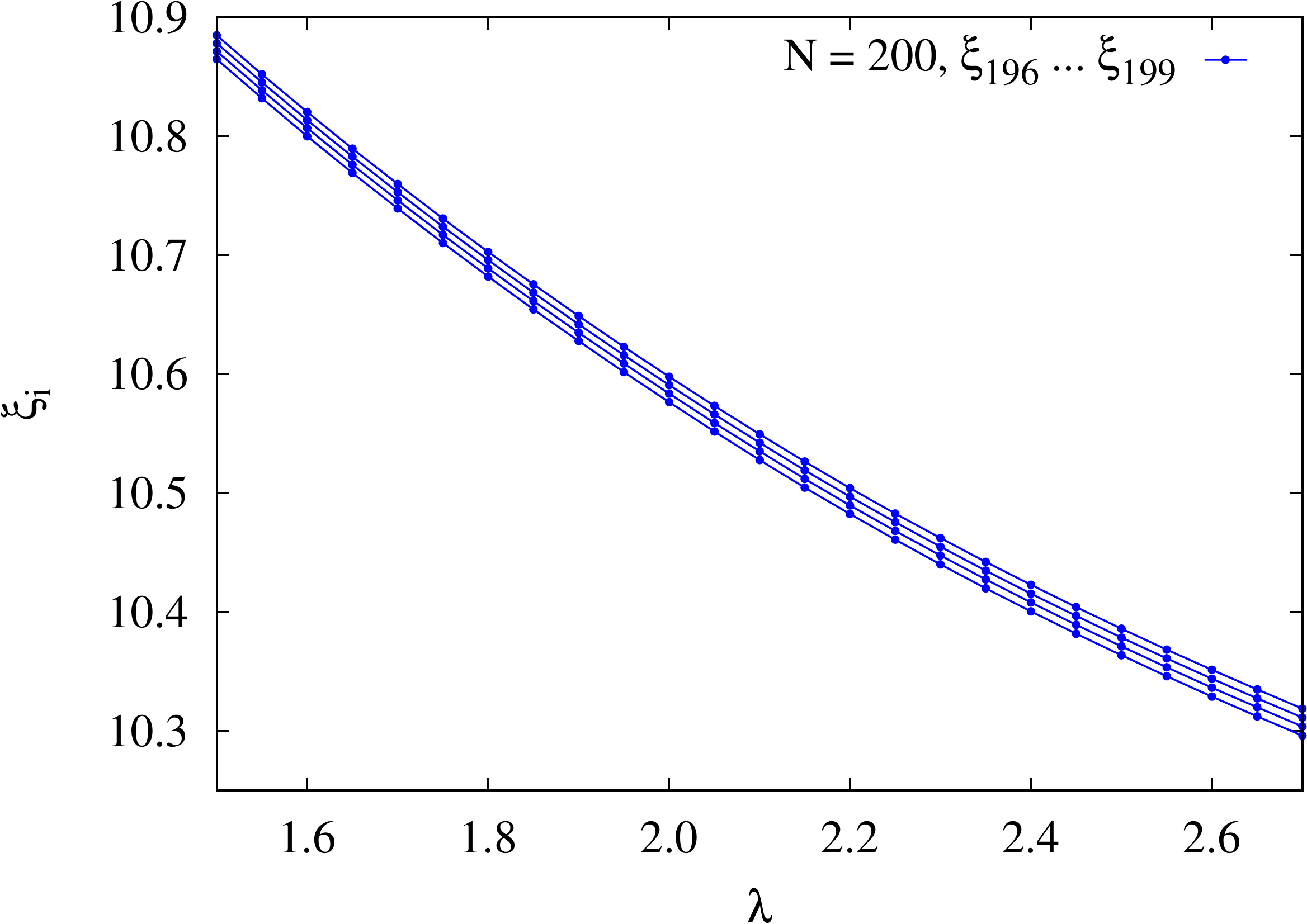}\\
  \caption{Spectral properties of the Hessian matrix $H_0$ for the saddle point with $m = 0$ in the strong-coupling phase. On the left: logs of the lowest eigenvalue $\xi_0$ and the spacing between higher eigenvalues for $N = 200$ compared with the strong-coupling instanton action \iftoggle{arxiv}{(\ref{instanton_action_sc})}{ (Eq. (5) in the main text)}. On the right: $4$ highest eigenvalues of $H_0$.}
  \label{fig:hessian_spectral_flow}
\end{figure*}

In the saddle-point approximation, the Hessian matrix determines the next-to-leading order correction $\det{H}^{-1/2} = e^{-1/2 \tr \ln H}$ to the path integral originating from the Gaussian fluctuations around saddle points. It turns out that despite all the interesting properties of the eigenspectrum of the Hessian matrix, the full determinant obtained by multiplying all $N$ eigenvalues appears to be a rather smooth, almost constant function. In particular, it seems that the exponentially small contribution of the lowest eigenvalue is compensated in some way by the product of higher eigenvalues.

In order to illustrate the effect of Hessian determinant, on Fig.~\ref{fig:ReS_and_hessians} on the left we plot the Gaussian correction $-1/2 \tr \ln H$ to the saddle-point free energy and compare it with the leading contribution from the saddle action $S_0$ at $N = 30$ and $N = 40$. We also plot the analytic result for the planar free energy $E_0\lr{\lambda}$ \cite{Gross:80:1,Wadia:12:1}, which corresponds to $S_0/N^2$ at $N = \infty$, in order to visualize $1/N$ corrections to $S_0$:
\begin{eqnarray}
\label{planar_free_energy}
 E_0\lr{\lambda} =
 \begin{cases}
  \frac{2}{\lambda} + \frac{1}{2} \log\lr{\frac{\lambda}{2}} - 3/4, & \lambda < 2 \\
  \frac{1}{\lambda^2}, & \lambda \geq 2 \\
 \end{cases}
\end{eqnarray}
One can see that as compared to the leading contribution, the Gaussian correction due to the Hessian is a very slowly changing function of $\lambda$. Interestingly, this correction is always negative, thus increasing the saddle point contribution.

Furthermore, on the right plot on Fig.~\ref{fig:ReS_and_hessians} we plot the difference of the actions of the $m=0$ and $m = 1$ (at weak coupling) or $m=2$ (at strong coupling) saddles, both with and without the Gaussian corrections $1/2 \tr\ln{H}$ included:
\begin{eqnarray}
\label{dS_with_hessians}
 \Delta S =
 \begin{cases}
  S_1 - S_0, & \lambda < 2 \\
  S_0 - S_2, & \lambda \geq 2 \\
 \end{cases} ,
 \quad
 \Delta H =
 \begin{cases}
  H_1 - H_0, & \lambda < 2 \\
  H_0 - H_2, & \lambda \geq 2 \\
 \end{cases}
 ,
\end{eqnarray}
where $H_m$ denote the Hessian matrices for the saddles with instanton number $m$. For comparison, we also plot the exact large-$N$ instanton actions \iftoggle{arxiv}{(\ref{instanton_action_wc}) and (\ref{instanton_action_sc})}{ in the weak- and the strong-coupling phases (see Eqs.~(4)~and~(5) in the main text)}. Again one can see that the Gaussian corrections do not result in any significant modifications of the saddle contributions. We also note that for finite values $N = 30$ and $N = 40$ used for these plots the transition from the weak-coupling to the strong-coupling regime is shifted towards larger $\lambda$ (of course, at finite $N$ this is no longer a phase transition). Unfortunately, we cannot calculate the Gaussian corrections at significantly larger $N$ and thus enable direct comparison with \iftoggle{arxiv}{(\ref{instanton_action_wc}) and (\ref{instanton_action_sc})}{the analytic expressions for the instanton action}, since the calculation of $\det{H}$ becomes numerically unstable at large $N$ and the comparison is anyway impossible due to numerical errors. Nevertheless, for $N = 30, \, 40$ it seems that the effect of the inclusion of the Hessian is very minor, and can be hardly distinguished from systematic errors due to finiteness of $N$. We thus conclude that the fluctuations around nontrivial saddles in the GWW model \iftoggle{arxiv}{(\ref{eq:Action})}{} should not change the exponential factors $e^{-m \, N \, S_I}$ in the trans-series.

\begin{figure}
  \centering
  \includegraphics[width=0.5\linewidth]{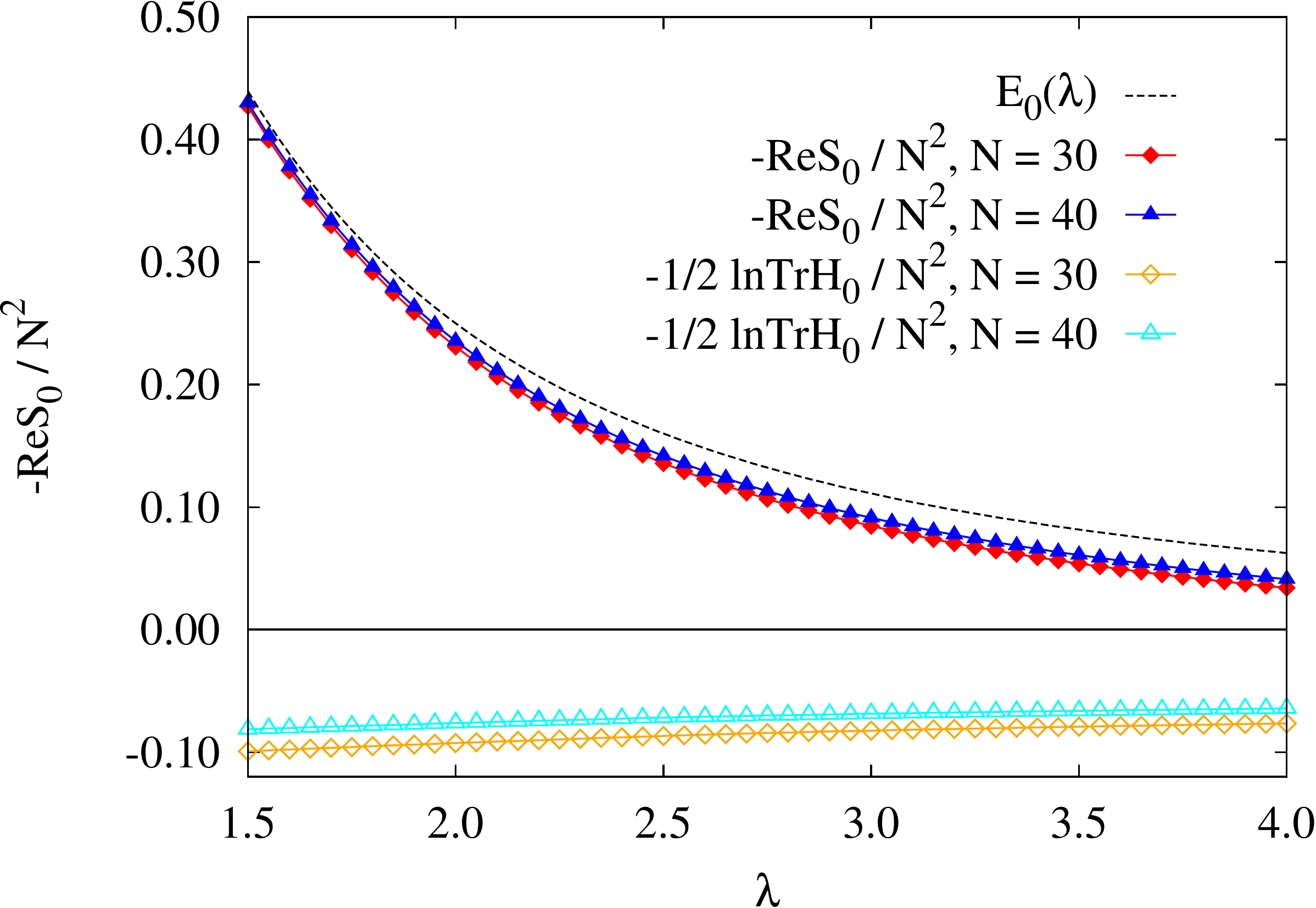}\includegraphics[width=0.5\linewidth]{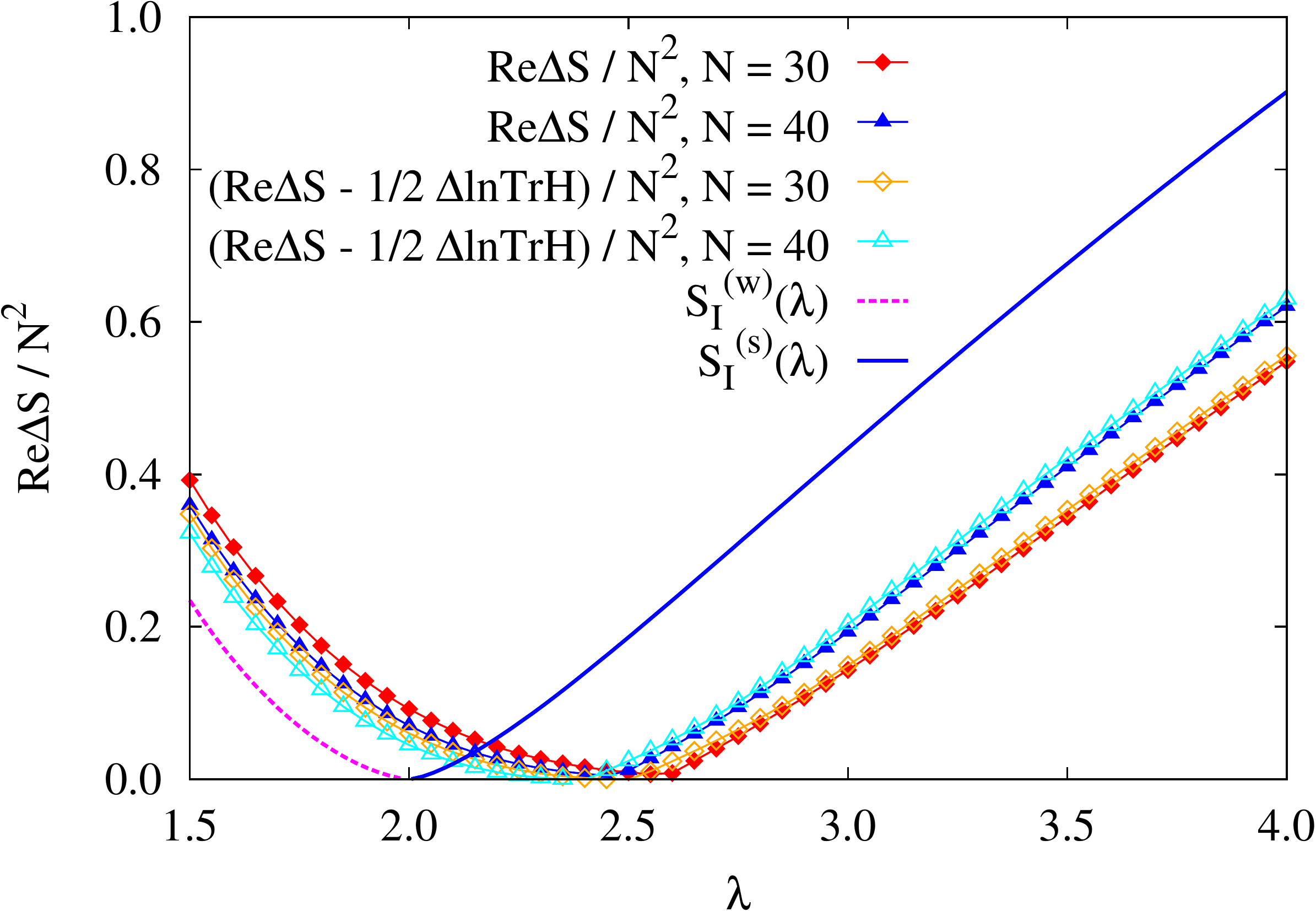}\\
  \caption{The effect of the Gaussian corrections $-1/2 \tr\ln H$ to the saddle action on the action of the vacuum saddle with $m = 0$ (on the left) and on the difference of the actions of $m=0$ and $m=1$ (in the weak-coupling phase) or $m=2$ (in the strong-coupling phase) saddles. We see that the effect of the Gaussian correction is very small and is hardly distinguishable from finite-$N$ corrections to the actions.}
  \label{fig:ReS_and_hessians}
\end{figure}

\end{widetext}
\end{document}